\documentclass[aps,twocolumn,prl,superscriptaddress]{revtex4}
\usepackage{graphicx}
\usepackage{amsmath}
\usepackage{amssymb}
\usepackage{epsfig}
\usepackage{wasysym}
\usepackage{bbm}
\renewcommand{\narrowtext}{\begin{multicols}{2} \global\columnwidth20.5pc}
\renewcommand{\v}[1]{{\bf #1}}

\def\be{\begin{eqnarray}}
\def\ee{\end{eqnarray}}
\newcommand{\nn}{\nonumber\\}
\newcommand{\Eq}[1]{Eq.~(\ref{#1})}
\newcommand{\Fig}[1]{Fig.~(\ref{#1})}

\newcommand{\nd}{{\vphantom{\dagger}}}

\begin{document}
\draft

\title{A Numerical Renormalization Group Study of the Superconducting and
Spin Density Wave Instabilities in a Two-band Model of MFeAsO$_{1-x}$F$_x$ Compounds}

\author{Fa Wang}
\affiliation{Department of Physics,University of California at Berkeley,
 Berkeley, CA 94720, USA}
\affiliation{Materials Sciences Division, Lawrence Berkeley National Laboratory,
 Berkeley, CA 94720, USA.}

\author{Hui Zhai}
\affiliation{Department of Physics,University of California at Berkeley,
 Berkeley, CA 94720, USA}
\affiliation{Materials Sciences Division, Lawrence Berkeley National Laboratory,
 Berkeley, CA 94720, USA.}

\author{Ying Ran}
\affiliation{Department of Physics,University of California at Berkeley,
 Berkeley, CA 94720, USA}
\affiliation{Materials Sciences Division, Lawrence Berkeley National Laboratory,
 Berkeley, CA 94720, USA.}

\author{Ashvin Vishwanath}
\affiliation{Department of Physics,University of California at Berkeley,
 Berkeley, CA 94720, USA}
\affiliation{Materials Sciences Division, Lawrence Berkeley National Laboratory,
 Berkeley, CA 94720, USA.}

\author{Dung-Hai Lee}
\affiliation{Department of Physics,University of California at Berkeley,
 Berkeley, CA 94720, USA}
\affiliation{Materials Sciences Division, Lawrence Berkeley National Laboratory,
 Berkeley, CA 94720, USA.}
\date{\today}

\begin{abstract}
We apply the fermion renormalization group method\cite{RG}, implemented numerically in Ref.\cite{Honerkamp}, to a two-band model of FeAs-based materials.  At half filling we find the $(\pi,0)$ or $(0,\pi)$ spin density wave order and a sub-dominant superconducting pairing tendency. Due to a topological reason, the spin density wave gap has nodes on the fermi surfaces. Away from half filling we find an unconventional $s$-wave and a sub-dominant
$d_{x^2-y^2}$ pairing instability. The former has $s$ symmetry around the hole fermi surface but exhibits  $s+d_{x^2-y^2}$ symmetry around the electron pockets where the 90 degree rotation is broken. The pairing mechanism is inter-pocket pair hopping. Interestingly, the same interaction also drives the antiferromagnetism.

\end{abstract}

\maketitle

Recently there is a flurry of interest in the the compound MFeAsO$_{1-x}$F$_x$\cite{Tc-Japan}. It has been shown that by varying the rare earth elements (M)\cite{Tc-2}, these materials can be made superconducting with T$_c$ as high as 55K\cite{Tc}. This has stimulated a flurry of interest in these materials. At the present time, preliminary experimental results have indicated that these materials
exhibit semimetallic antiferromagnetism at stoichiometry (i.e., $x=0$). Upon substituting O with F the antiferromagnetism diminishes while superconductivity appears\cite{Competing}.

\begin{figure}
\includegraphics[scale=0.4]{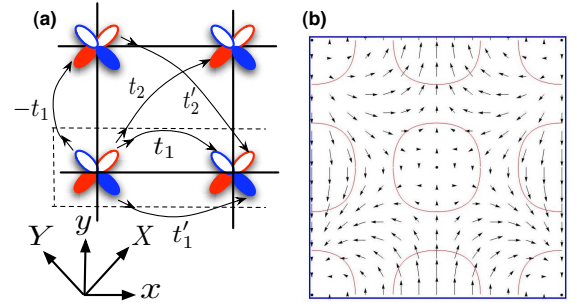}
\caption{(a) The in-plane projection of $d_{XZ}$ and $d_{YZ}$ orbitals and the four independent hopping parameters. (b) $\v b$ as a function of $\v k$. The red curves are the fermi surfaces.
}\label{fig:bz}
\end{figure}

Structurally MFeAsO$_{1-x}$F$_x$ can be viewed as As-Fe-As tri-layers separated by MO$_{1-x}$F$_x$ spacers. Early electronic structure calculations\cite{LDA} and angle-integrated photoemission\cite{feng} suggest that the carriers at the fermi energy are essentially Fe in character.  Thus, like many others, we will focus on a As-Fe-As tri-layer in the following discussions. To envision these tri-layers, imagine a square lattice of Fe. The As sit either above or below the center of the square plaquettes, and these two types of plaquette form a checkerboard. Each structural unit cell contains two Fe (dashed rectangle in \Fig{fig:bz}(a), and the basis vectors are $\v X=\hat{x}+\hat{y}$ and $\v Y=\hat{x}-\hat{y}$ where $\hat{x}$ and $\hat{y}$ are the basis vectors of the Fe square lattice (see \Fig{fig:bz}(a)). In the following the we shall refer to the reciprocal unit cell associated with the basis vector $\hat{x},\hat{y}$ and $\v X,\v Y$ as the ``unfolded'' and ``folded'' Brill!
 ouin zone (BZ) respectively.

Aside from the semi-metalicity, there are other important differences between the antiferromagnetism in this material and that of the cuprates. First, in the unfolded BZ the magnetic ordering wavector is either $(0,\pi)$ or $(\pi,0)$\cite{neutron} instead of the usual $(\pi,\pi)$. Second, the ordering moment is quite small ($\sim 0.25\mu_B$)\cite{neutron} compared to that of the cuprates. For the superconducting state, penetration depth\cite{penetration}, H$_{c2}$\cite{hc2}, $1/T_1$ of nuclear spins\cite{nmr}, $\mu$SR\cite{musr}, and point tunneling measurements\cite{tunneling} all indicate the presence of line nodes in the superconducting gap.
However, more experiments, in particular those done on single crystals, will be necessary to check the above conclusions.

On the theoretical side, several electronic structure calculations suggest the presence of hole and electron pockets at fermi energy. It is suggested that the near nesting of these pockets is responsible for the antiferromagnetism at $x=0$\cite{LDA,LDA-2,nesting}. Estimate of the strength of local coulomb interaction 
suggests that this system is on the border between strong and weak correlation\cite{kotliar}. Although LDA type calculations suggest all five orbitals of Fe contribute to states at the fermi energy\cite{LDA}, this was simplified recently where only two out of the five bands are kept while preserving the electron and hole pockets\cite{TaoLi}.  Based on different degree of simplification of the electronic structure, and different approximate treatment of the electronic correlation, a number of groups have studied the superconducting
pairing instability of these material\cite{sc}. In addition there are a couple of recent attempts to use
the point group symmetry to narrow down the pairing symmetries\cite{symmetry}. The wide spread in
pairing symmetry concluded from these studies call for a more unbiased assessment of the
pairing instability.

In this paper we study a two band model with Hubbard-like and Hund interactions. We study the possibility of electronically induced pairing by performing one-loop renomalization group (NRG) calculation\cite{RG}. This numerical version of this method (NRG) was applied to the cuprates by Honerkamp et al\cite{Honerkamp}. It was shown that in the framework of one-band Hubbard model, interactions which promote $d_{x^2-y^2}$ pairing and $(\pi,\pi)$ antiferromagnetic order are generated at low energies. We believe that owing to the weaker correlation, this method is much better suited for MFeAsO$_{1-x}$F$_x$. The NRG results for five band model \cite{LDA} will be the subject of upcoming publication\cite{fawang}.

{\it Model Hamiltonian -} As Ref.~\cite{TaoLi} we simplify the electronic structure of the As-Fe-As tri-layter by keeping only two Fe orbitals: 3d$_{xz}$ and 3d$_{yz}$. The Arsenic are viewed as merely mediating hopping between these orbitals. Due to the relative orientation of the Fe and As it is more convenient to use as basis 3d$_{XZ}$ and 3d$_{YZ}$ where $x,y$ and $X,Y$ are shown in \Fig{fig:bz}(a). Our tight binding model
include nearest-neighbor and next-nearest-neighbor hoppings.
From symmetry considerations there are four
independent hopping parameters $t_1,t_1',t_2,t_2'$ as shown in \Fig{fig:bz}(a). Among them $t_1'$ is due to the direct overlap of two neighboring Fe orbitals. All the rest three parameters describe hopping mediated by As.
In the following we shall label the Fe orbitals 3d$_{XZ}$ and 3d$_{YZ}$ as
$a=1,2$.
The tight-binding Hamiltonian in the unfolded BZ reads
\be
 &&\hat{H}_0=\sum_{\v{k},s}\sum_{a,b=1}^{2}c_{a\v{k} s}^\dagger K_{ab}^\nd(\v{k})c_{b\v{k} s}^\nd =\sum_{\v k,s}\sum_{n=1}^2\epsilon_n^\nd(\v k)\psi^\dagger_{n\v{k} s}\psi_{n\v{k} s}^\nd \nn
&&K(\v k)=\alpha(\v k)I+b_x(\v k)\tau_x+b_z(\v k)\tau _z,
\label{h0}\ee
where $c_{a\v{k} s}$ annihilates a spin $s$ electron in orbital $a$ and momentum $\v{k}$, and $\psi_{n\v k s}$ is the band annihilation operator. The $I,\tau_x,\tau_y$ in  \Eq{h0} are the $2\times 2$ identity matrix and Pauli matrices respectively. They act  on the orbital $(d_{XZ},d_{YZ})$ space, and $\alpha(\v k)=\mu +2 t_1' \cos k_y+2\cos k_x[t_1'+(t_2+t_2')\cos k_y]$, $b_x(\v k)=2t_1(\cos k_x-\cos k_y)$, $b_z(\v k)=-2\left(t_2-t_2'\right)\sin k_x\sin k_y$, and  $\epsilon_{1,2}(\v k)=\alpha(\v k)\mp \sqrt{b_x(\v k)^2+b_z(\v k)^2}$.
We have checked that when acted upon by the element ($g$) of the point group ($C_{4v}$) the band operator
$\psi_{a\v{k} s}\rightarrow \eta_g\psi_{a g\v{k} s}$ where $\eta_g=\pm 1$.
After turning on a proper chemical potential $\mu$ we get two hole pockets
around $\v k=(0,0)$ and $(\pi,\pi)$ and two electron pockets at
$(0,\pi)$ and $(\pi,0)$. In the rest of the paper we shall use the following values for the hopping parameters $t_1=0.38 eV, t_2=0.57 eV, t_2'=0$. The value of $t_1'$ critically determines the superconducting gap function, and will be discussed later.

Now we consider local interactions
including intra-orbital and inter-orbital Coulomb interaction $U_1$,
and  $U_2$, Hund's coupling $J_H$ and the pair hopping term. When summed together they give
\begin{eqnarray}
&& \hat{H}_{\rm int}=\sum_{i}\{U_1\sum_{a=1}^2
 n_{i,a,\uparrow}n_{i,a,\downarrow}+U_2n_{i,1} n_{i,2} \nonumber\\
&& +
 J_H[\sum_{s,s'}c_{i1 s}^\dagger c_{i2 s'}^\dagger c_{i1 s'}^\nd c_{i2 s}+ (c^\dagger_{i1\uparrow}c^\dagger_{i1\downarrow}c_{i2\downarrow}c_{i2\uparrow}+h.c.) ]\}.\nonumber
 \end{eqnarray}
Here $i$ labels the unit cell, $s,s'=\uparrow,\downarrow$, and
$n_{i,a}=n_{i,a,\uparrow}+n_{i,a,\downarrow}$ is the number operator associated with orbital $a$. The total Hamiltonian
$\hat{H}=\hat{H}_0+\hat{H}_{\rm int}$ is the starting point of our study.

\begin{figure}
\includegraphics[scale=0.95]{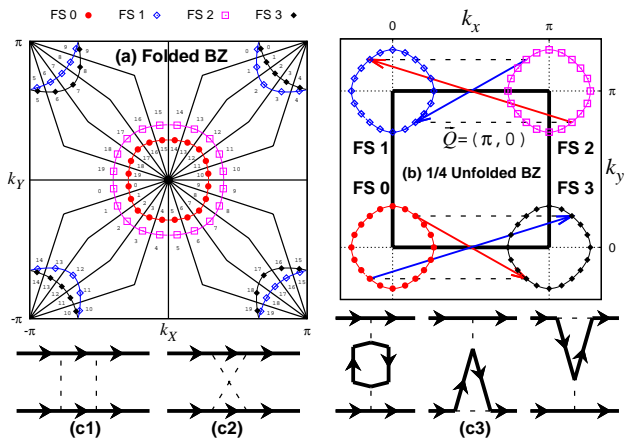}
\caption{ (a) The division of the first BZ into N (=20) patches. The color curves denote the fermi surface($t_1'=-0.05$). (b) The strongest inter-pocket pair tunneling(for $t_1'=0$). As indicated by the red and blue arrows, a $\pm \v q$ pair on one pocket fermi surface are scattered to a $\pm \v k$ pair on another pocket. Interestingly, the same scattering also drives the antiferromagnetism. This can be seen by noting that the momentum transfer between the incoming spin up(red) and outgoing spin down(blue) electrons is exactly the nesting wavevector $(\pi,0)$ or $(0,\pi)$ as indicated by the dashed line. (c1)-(c3) The summed Feynman diagrams. Here the arrowed solid denote Green's function and the dashed lines represent the renormalized interactions.
}\label{fig:NRG}
\end{figure}

We have checked that the bare $\hat{H}$ has no superconducting instability for realistic interaction parameters.
Thus we decide to perform a numerical renormalization group (NRG) calculation where high energy electronic excitations are recursively integrated out. The hope is that the effective interaction generated at low energy would show the sign of superconducting instability. in the following we present the result of such a calculation. Technical details of the NRG can be found in Ref.~\cite{Honerkamp}. In brief, we divide the first BZ into N patches, and at each renormalization iteration we sum over the five one-loop Feynman diagrams labeled (c1)-(c3) in \Fig{fig:NRG}. The essential complication in the present study is the presence of two different bands and four disjoint pieces of fermi surface.  It turns out that it is easier to work with the folded BZ. This is because folding doubles the number of bands so that each band has only one fermi surface (\Fig{fig:NRG}(a)) and the BZ patching scheme of Ref.\cite{Honerkamp} can be directly applied. Because of the existence of four band!
 s, we need to keep track of the band indices as well as the momenta in the interaction vertex.  Thus we need to compute $256\times N^3$ different interaction vertices at each renormalization step in contrast to $N^3$ in the single band case\cite{Honerkamp}. Although our RG calculation is performed with the folded BZ we shall present our results using the unfolded BZ for simplicity.

{\it The antiferromagnetism and superconducting pairing tendency at half filling.}
At stoichiometry (x=0) the two bands derived from the $d_{XZ}$ and $d_{YZ}$ orbitals accommodate 2 electrons per unit cell. For $t_1'=0$ the electron and hole pockets are perfectly nested by $(0,\pi),(\pi,0)$ in the unfolded BZ. 
In a theory like ours, the antiferromagnetism is due to the the above nesting. For $t_1'\ne 0$ the nesting is imperfect. Hence one might expect a full spin density wave (SDW) gap for $t_1'=0$ and a partially gapped fermi surface for $t_1'\ne 0$. However, as we will show in the following, even for the former case there are nodes in the SDW gap, and the reason is topological.

\begin{figure}[htcp]
\includegraphics[scale=.8]{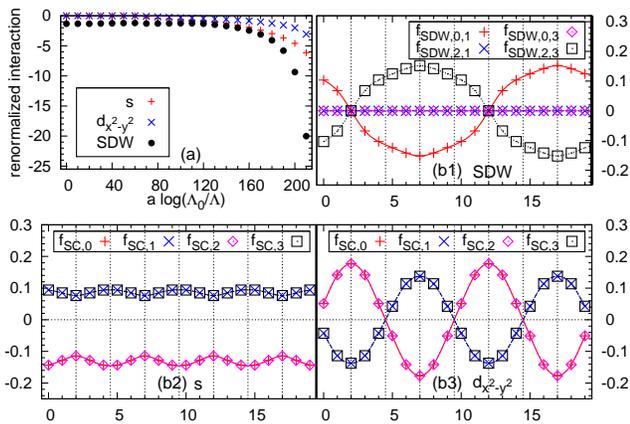}
\caption{(a)The N=20 RG evolution of the scattering amplitudes in the SDW and two different types of superconducting pairing channels as a function of  $a\ln(\Lambda_0/\Lambda)$. The parameter $t_1'$ is set to zero. (b1) The SDW form factor is plotted as the momentum moves around the $(0,0)$ or $(\pi,\pi)$ fermi surface. The vertical dash lines are high symmetry directions. The curves are obtained by interpolating the data points. Here the flat (zero) form factor are associated with the fermi surfaces that are not nested by the ordering wavevector $(\pi,0)$ or $(0,\pi)$. (b2,b3) The form factor of the two most prominent superconducting pairing. The most favorable pairing symmetry is u-$s$-wave and the next one is $d_{x^2-y^2}$-wave. }\label{flow}
\end{figure}

First, we say a few words about how the results are obtained. In the calculation we compute the renormalized interaction vertex function $V(\v k_1,a;\v k_2,b;\v k_3,c;d)$ for two incoming electrons with opposite spins. Here $a,b,c,d=1,..,4$ are the band indices, and $\v k_1,..,\v k_3$ are momenta on the Fermi surfaces. From this vertex function we can extract the renormalized interaction in the Cooper or SDW channels as follows. For the singlet/triplet Copper channel $V^{SC}_{s,t}(\v k,a;\v p,b)=V(\v k,a;-\v k,a;\v p,b;b)\pm V(-\v k,a;\v k,a;\v p,b;b)$; for the SDW channel $V^{SDW}(\v k,a,d;\v p,c,b)=-2 V(\v k,a;\v p+\v Q,b;\v p,c;d)$, where $\v Q$ is the ordering wavevecctor, band indices $b=c\pm 1\mod 4$ and $d=a\pm 1\mod 4$.
Since $\v k$ and $\v p$ only takes $N$ discrete values, we can treat $V^{SC}_{s,t}$ and $V^{SDW}$ as matrices. The few lowest eigenvalues as a function of the RG evolution (panel (a)) and their final associated eigenvectors (panel (b)) of these matrices are plotted in \Fig{flow}.

In panel (a) of \Fig{flow} we show the N=20 RG evolution of the scattering amplitudes in the SDW and two different types of superconducting pairing channels as a function of  $a\ln(\Lambda_0/\Lambda)$. Here $a=-1/\ln(0.97)$, $\Lambda_0$ is the initial and $\Lambda$ is the running energy cutoff.
The bare interaction parameters used are $U_1=4.0,\,U_2=2.5,\,J_H=0.7$ eV. Clearly as $\Lambda_0/\Lambda$ increases the interaction that drives SDW (black dots) grows in magnitude the fastest. The form factor $f_{SDW}(\v k)$ associated with the SDW order
($\hat{\Delta}_{SDW,a,b}=\sum_{\v k} f^\nd_{SDW,a,b}(\v k) \psi^\dagger_{b\v k+(\pi,\pi)\uparrow} \psi^\nd_{a\v k\downarrow}$, $b=a\pm 1\mod 4$) is shown in panel (b1) of this figure.  Interestingly despite the perfect nesting of the fermi surfaces (see Fig.\ref{fig:NRG}(b)) there are two nodes !

To understand the origin of these nodes we observe that the top of the hole pockets (situated at $\v k=(0,0)$ and $(\pi,\pi)$) is doubly degenerate. According to Berry \cite{berry} the band eigenfunctions must exhibit non-trivial phase as $\v k$ moves around these degenerate $\v k$ points.
Omitting the identity term, the $K(\v k)$ in \Eq{h0}, $K(\v k)=b_z(\v k)\tau_z+b_x(\v k)\tau_x$, is that of a spin 1/2 in a $\v k$-dependent magnetic field. A plot of $\v b$ as a function of $\v k$ for $t_1'=t_2'=0$ is given in \Fig{fig:bz}(b). Clearly, as $\v k$ circles around $(0,0)$ or $(\pi,\pi)$ the direction of $\v b$ winds twice around the unit circle. This ``double-winding'' explains the fact that the degeneracy of the band dispersion is lifted ``quadratically'' as $\v k$ deviates from $(0,0)$ or $(\pi,\pi)$. On the contrary, the bottom of electron pockets around $(0,\pi)$ and $(\pi,0)$ are non-degenerate, and $\v b$ exhibits no winding around them. Now let us consider switching on a SDW order parameter to nest, say, the fermi surfaces around $(0,0)$ and $(\pi,0)$. Let $\v q$ be the momentum around the $(0,0)$-fermi surface, and  $|\psi(\v q)\rangle$ and
$|\psi(\v q+(\pi,0))\rangle$ be, respectively, the band eigenstates associated with the two fermi surfaces. The
following matrix element
$\Delta(\v q)=\langle\psi(\v q)|M(\v q)|\psi(\v q+(\pi,0))\rangle$ determines the SDW gap. Here $M(\v q)$ is a $2\times 2$ matrix acting in the orbital space (here we have assumed that after choosing a spin quantization axis, the ordered moments lies, say, in the $\pm$ x-direction). From \Fig{fig:bz}(b) one can see that
$H(\v q)=H(-\v q)$ and $H((\pi,0)+\v q)=H((\pi,0)-\v q)$. 
However, due to the double-winding behavior of $\v b$ around $(0,0)$ there is a non-trivial Berry phase after $\v k$ made a half circle around the origin, i.e., $|\psi(-\v q)\rangle=-|\psi(\v q)\rangle$. On the other hand the no-winding of $\v b$ around $(\pi,0)$ implies $|\psi(-\v q+(\pi,0))\rangle=|\psi(\v q+(\pi,0))\rangle$. Consequently if $M(\v q)$ is inversion symmetric, i.e., $M(\v q)=M(-\v q)$, we have $\Delta(\v q)=-\Delta(-\v q).$ Under the assumption that the magnetically ordered phase preserves time reversal plus a spatial translation (hence $\Delta(\v q)$ is real), this implies the gap function must change sign twice as $\v q$ moves around the fermi surface. Hence there must be two diametrically opposite nodes.
Explicit mean-field calculation\cite{ying} using the bare $H$ shows that nodes are situated at the intersection between the ordering wave vector and the $(0,0)$-fermi surface. This agrees with the form factor of \Fig{flow}(b1).

Next, we come to superconducting pairing. As shown in \Fig{flow}(a) even for half-filling there are growing interaction that drives superconductivity! However, these interaction are sub-dominant compared with the interaction that promotes antiferromagnetism. Our result shows that the two most favorable pairing symmetry are an unconventional $s$ (u-$s$) and $d_{x^2-y^2}$ like. The u-$s$-wave pairing has $s$ symmetry around the hole fermi surface but exhibits $s+d_{x^2-y^2}$ symmetry around the electron fermi surface (where the 90 degree rotation symmetry is broken)\cite{LDA-2,symmetry}. It turns out that depending on the value of $t_1'$ it is possible for the gap function to have nodes on the electron fermi surface. We shall return to this point shortly. The form factors $f_{SC,a}(\v k)$ ($\Delta_{SC,a}= f_{SC,a}(\v k)[\psi_{a\uparrow}(\v k)\psi_{a\downarrow}(-\v k)-\uparrow\leftrightarrow\downarrow])$
of these pairing symmetry are shown in \Fig{flow}(b2,b3). {\it Most significantly, from our calculation the pairing mechanism can be unambiguously determined - the pairing are all driven by the inter-pocket pair tunneling}\cite{tband}. An example of the strongest such process is shown in \Fig{fig:NRG}(b). As indicated by the red and blue arrows, a $\pm \v q$ pair on one fermi surface are scattered to a $\pm \v k$ pair on another. Interestingly, the same scattering also drives the antiferromagnetism. This can be seen by noting that the momentum transfer between the incoming spin up and outgoing spin down electrons is exactly the nesting wavevector $(\pi,0)$ or $(0,\pi)$ as indicated by the dashed line. Thus the same interaction also drives antiferromagnetism! One might ask would't pairing and SDW interaction require opposite sign? No, for the inter band pairing interaction either sign will do. This is
because the pairing order parameters can choose opposite signs on the two bands thus benefit from the positive interaction.
At half-filling  when the nesting is sufficiently good (e.g., when $t_1'$ is absent), antiferromagnetism overwhelms the superconducting instability. We propose the reason weak superconductivity observed in the stoichiometric compound LaFePO is because As$\leftrightarrow$P replacement damages nesting hence allow superconductivity to prevail in the competition with antiferromagnetism.

\begin{figure}
\includegraphics[scale=1]{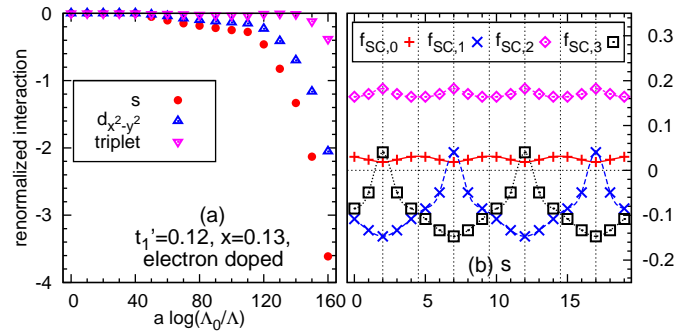}
\caption{(a) The N=20 RG evolution of the scattering amplitude associated with the two most favorable singlet channel (u-$s$ and $d_{x^2-y^2}$) and the top triplet channel. (b) The form factors of u-$s$.  Note that it changes sign on the electron pocket.}\label{flow1}
\end{figure}

{\it The superconducting pairing away from half filling}
In panel (a) and (b) of \Fig{flow1} we show the N=20 RG evolution of the two most favorable singlet and the best triplet superconducting scattering amplitudes for $x=0.13$. The $t_1'$ parameter we used here is $0.12$ eV, and the bare interaction parameters are $U_1=4.0,U_2=2.5,J_H=0.7$ eV. Due to the removal of nesting, the antiferromagnetic scattering (not shown) is no longer dominant. Like half-filling, the most favorable pairing symmetry is the singlet u-$s$. The pairing mechanism is the inter-pocket pair hopping shown in \Fig{fig:NRG}(b). At $t_1'=0.12$ eV the form factor changes sign as shown in \Fig{flow1}(b). As a result {\it the superconducting gap has nodes on the electron pocket}. For all parameters we have studied, the triplet pairing channel is never favored. In \Fig{flow1}(a) we show the RG evolution of the best triplet pairing amplitude, and it never becomes competitive with the u-$s$ channel. We emphasize that while the u-$s$ gap function always show a full gap on the hole pockets, it can be gapped or gapless on the electron pocket depending on the value of $t_1'$. For the parameter range we have studied a systematic trend is clearly visible: larger $t_1'$ makes u-$s$ gapless. {\it Under the assumption that our two band model describes the
band structure adequately}, we propose that this is the superconducting pairing that has been observed experimentally.
\\

Acknowledgement: We thank Henry Fu for most helpful discussions.
DHL was supported by DOE grant number DE-AC02-05CH11231. AV was supported by LBNL DOE-504108.


\begin{thebibliography}{99}
\bibitem{RG}  R. Shankar, Rev. Mod. Phys. {\bf 66}, 129 (1994); J. Polchinski, Proceedings of 1992 Theoretical Advanced Studies Institute in Elementary Particle Physics, edited by J. Harvey and J. Polchinski, World Scientific, Singapore 1993.
\bibitem{Honerkamp}
C. Honerkamp, {\it et.al.} Phys. Rev. B. {\bf 63}, 035109 (2001)
\bibitem{Tc-Japan}
Y. Kamihara, JACS, {\bf 128}, 10012 (2006) and Y. Kamihara, JACS, {\bf 130}, 3296 (2008)
\bibitem{Tc-2}
G. F. Chen, {\it et.al.} arxiv:0803.3790;
X. H. Chen, {\it et.al.} arXiv: 0803.3603;
P. Cheng, {\it et.al} arXiv: 0804.0835 and Hai-Hu Wen {\it et.al.} Europhys. Lett. {\bf 82} 17009 (2008)

\bibitem{Tc}
J. Yang {\it et.al.} arXiv: 0804.3727 and Zhi-An Ren {\it et.al.}, arXiv: 0804.2053

\bibitem{Competing}
J. Dong {\it et.al.} arXiv: 0803.3426, R. H. Liu, {\it et.al.} 0804.2105 and Y. Qiu {\it et.al.} arXiv: 0805.1062;


\bibitem{LDA}
C. Cao, P. J. Hirschfeld, H.  P. Cheng, arXiv:0803.3236; F. Ma, Z. Y. Lu arXiv:0803.3286, D.J. Singh, M.H. Du arXiv:0803.0429; K. Kuroki, {\it et.al.} arXiv:0803.3325

\bibitem{LDA-2}
I. I. Mazin, {\it et.al.} arXiv: 0803.2740

\bibitem{nesting}
V.  Cvetkovic and Z. Tesanovic, arXiv: 0804.4678

\bibitem{feng}
H. W. Ou {\it et.al}, arXiv: 0803.4328
\bibitem{neutron}
C. Cruzar, {\it et.al.} arXiv:0804.0795; H.-H. Klauss {\it et.al.} arXiv:0805.0264
S. Kitao {\it et.al.} arXiv:0805.0041

\bibitem{penetration}
K. Ahilan, {\it et.al.}, arXiv:0804.4026

\bibitem{hc2}
C. Ren {\it et.al.} arXiv:0804.1726; F. Hunte {\it et.al.} arXiv:0804.0485

\bibitem{nmr}
Y. Nakai, {\it et.al.} arXiv: 0804.4765

\bibitem{musr}
H. Luetkens {\it et.al.} arXiv:0804.3115
\bibitem{tunneling}
L. Shan {\it et.al.} arXiv:0803.2405
\bibitem{kotliar}
K. Haule, J. H. Shim, G. Kotliar arXiv: 0803.1279

\bibitem{TaoLi}
T. Li, arXiv: 0804.0536;
S. Raghu {\it et.al.} arXiv: 0804.1113;
Q. Han, Y. Chen, Z. D. Wang, Europhys. Lett. {\bf 82} 37007 (2008)


\bibitem{sc}
X. Dai {\it et.al.} arXiv: 0803.3982, P. A. Lee, X. G. Wen, arXiv:0804.1739;
X. L.  Qi {\it et.al.} arXiv:0804.4332;
Z.J. Yao, J. X. Li, Z. D. Wang arXiv:0804.4166; Q. Si, E. Abrahams arXiv:0804.2480; G. Baskaran, arXiv: 0804.1341; Z. Y. Weng, arXiv: 0804.3228; F. J. Ma, Z.-Y Lu, T. Xiang, arXiv:0804.3370 and K. Seo, B. A. Bernevig and J. P. Hu, arXiv:0805.2958.

\bibitem{symmetry}
Z. H. Wang {\it et.al.} arXiv: 0805.0736, Y. Wan, Q. H. Wang  arXiv: 0805.0923

\bibitem{fawang} Fa Wang {\it et al}, arXiv:0807.0498.

\bibitem{berry} M. Berry, Proc. R. Soc. London, A {\bf392}, 45 (1984)

\bibitem{ying} Ying Ran {\it et al}, arXiv:0805.3535.

\bibitem{tband} H. Suhl, B.T. Matthias, L.R. Walker, Phys. Rev. Lett. {\bf 12}, 552 (1959).



\end{thebibliography}
\end{document}